# Low-threshold InP quantum dot and InGaP quantum well visible lasers on silicon (001)

**PANKUL DHINGRA,**[1,2] **PATRICK SU,**[1,2] **BRIAN D. LI,**[1,2] **RYAN D. HOOL,**[2,3] **AARON J. MUHOWSKI,**[4] **MIJUNG KIM,**[1,2] **DANIEL WASSERMAN,**[4] **JOHN DALLESASSE,**[1,2] **AND MINJOO LARRY LEE**[1,2,*]

[1]Department of Electrical and Computer Engineering, University of Illinois at Urbana-Champaign, 306 N Wright Street, Urbana, IL, 61801, USA

[2]Holonyak Micro and Nanotechnology Laboratory, University of Illinois at Urbana-Champaign, 208 N Wright Street, Urbana, IL, 61801, USA

[3]Department of Materials Science and Engineering, University of Illinois at Urbana-Champaign, 1304 W Green Street, Urbana, IL, 61801, USA

[4]Department of Electrical and Computer Engineering, University of Texas at Austin, 2501 Speedway, Austin, Texas, 78712, USA

*Corresponding author: mllee@illinois.edu

**Monolithically combining silicon nitride (SiN$_x$) photonics technology with III-V active devices could open a broad range of on-chip applications spanning a wide wavelength range of ~ 400 – 4000 nm. With the development of nitride, arsenide, and antimonide lasers based on quantum well (QW) and quantum dot (QD) active regions, the wavelength palette of integrated III-V lasers on Si currently spans from 400 nm to 11 μm with a crucial gap in the red-wavelength regime of 630 – 750 nm. Here, we demonstrate the first red In$_{0.6}$Ga$_{0.4}$P QW and far-red InP QD lasers monolithically grown on CMOS compatible Si (001) substrates with continuous-wave operation at room temperature. A low-threshold current density of 550 A/cm$^2$ and 690 A/cm$^2$ with emission at 680-730 nm was achieved for QW and QD lasers on Si, respectively. This work takes the first vital step towards integration of visible red lasers on Si allowing the utilization of integrated photonics for applications including biophotonic sensing, quantum computing, and near-eye displays.**

The primary application for Si photonics technology to date is integrated photonic transceivers for telecommunications where off-chip or hybrid-integrated InP-based lasers emitting in the C- or O- bands (~1.3-1.6 μm) serve as the light source [1]. Leveraging the CMOS foundry, Si photonics now enables an increasing number of applications including mapping and navigation [2], spectroscopy [3], and quantum communication [4]. Applications that rely on visible lasers, such as biosensing [5], atomic clocks [6] and spatial mapping [2] could greatly benefit from the ability to generate, guide and sense light on a chip [7]. As another example, integrated photonics could help overcome the limitations of free-space optics for trapped-ion quantum computing relying on 674 nm lasers to drive transitions in $^{88}$Sr+ ion qubits [8]. Low-loss SiN$_x$ waveguide technology [9] is a key enabler for visible photonics chips, but efforts to integrate visible sources on Si are currently lagging behind.

Despite elevated threading dislocation density (TDD) values of ~10$^6$-10$^8$ cm$^{-2}$ [10], recent efforts have yielded a wide range of monolithic lasers on Si utilizing GaAs-, InP-, and GaSb-based active regions. For example, InAs QD lasers on GaAs/Si (001) emitting at 1.3 μm, have been demonstrated with threshold current density (J$_{th}$) values as low as 62.5 A/cm$^2$ for room temperature (RT), continuous-wave (CW) operation [11]; J$_{th}$ = 100-300 A/cm$^2$ is more typical for InAs QD lasers on Si [12, 13]. In addition, Shang et al. demonstrated an extrapolated operation lifetime of 22 years with a constant current stress at 80°C by preventing the formation of misfit dislocations in the InAs QD active region [12-14]. As another example, InGaAsP multiple quantum well (MQW) lasers on InP/Si (001) emitting at 1.55 μm with CW, RT operation and J$_{th}$ = 1-3 kA/cm$^2$ have also been reported [15, 16]. Finally, GaSb-based QW lasers emitting at 2.3 μm were recently demonstrated using on-axis Si substrates with J$_{th}$ = 200-300 A/cm$^2$ and CW operation up to 80°C [17, 18]; further references on epitaxial III-V lasers on Si can be found in recent review articles [10, 19, 20]. Despite the impressive development of near- and shortwave-infrared lasers on Si based on both QD and QW active regions, there are no reports for electrically injected red lasers on on-axis Si. Development of monolithic visible red lasers on Si would fill a crucial gap in the wavelength palette of integrated lasers on Si, which already spans from 400 nm [21] to 11 μm [22].

In$_x$Ga$_{1-x}$P QWs and InP QDs are a versatile platform for high-efficiency diode lasers emitting in the 630-800 nm wavelength regime. InP QD lasers on GaAs emitting in the 680-750 nm [23, 24] wavelength range have shown pulsed $J_{th}$ values as low as 190 A/cm$^2$ [24] and high-power output > 150 mW [25]. Compressively strained In$_x$Ga$_{1-x}$P (x > 0.49) QW lasers on GaAs are even more mature and have demonstrated watt-class power output [26] with $J_{th}$ of 295 A/cm$^2$ [27]. Kwon et al. demonstrated the first In$_{0.58}$Ga$_{0.42}$P QW lasers on 6° offcut Si (001) using a 10 μm-thick SiGe buffer with TDD = $2 \times 10^6$ cm$^{-2}$ [28]. Despite the well-controlled TDD, the lasers only operated pulsed with a high $J_{th}$ of 1.65 kA/cm$^2$, and moreover, the use of offcut Si substrates renders such devices incompatible with Si photonics foundries [28]. More recently, Luo et al. demonstrated optically pumped InP QD microdisk lasers on Si (001) [29]. Despite the excellent performance of phosphide-based visible lasers on GaAs, further work is required to monolithically integrate visible lasers and optical amplifiers on Si (001).

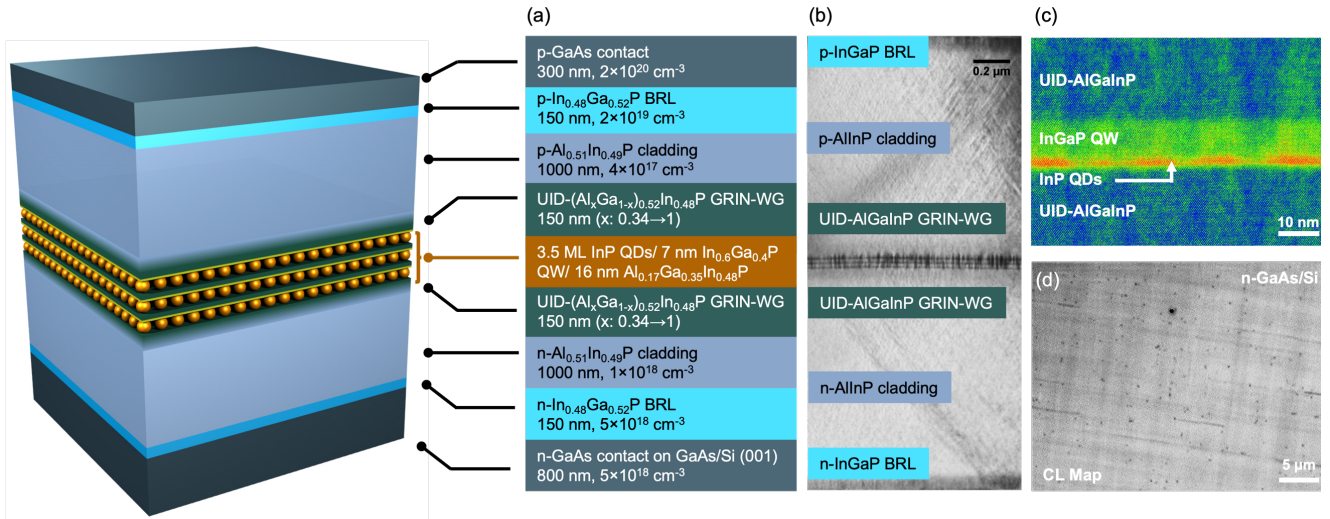

**Fig. 1.** (a) Growth schematic of InP MQD laser on GaAs/Si; the SQW laser is similar except without QDs and only one repeat of the active region (b) BF-TEM image of InP MQD laser on GaAs/Si showing strain contrast around the InP QDs. (c) False-colored high-resolution HAADF-STEM image of a single InP QDWELL layer showing composition contrast from individual QDs. (d) CL map of n-GaAs/Si showing TDD of $\sim 1 \times 10^7$ cm$^{-2}$ with dark spots correlating to threading dislocations.

To date, no electrically injected red laser has been demonstrated on exact Si (001), preventing visible integrated photonics from fully leveraging advances in high-performance SiN$_x$ passive optical components and Si photodetectors. In this article, we demonstrate the first visible In$_{0.6}$Ga$_{0.4}$P QW and InP QD lasers monolithically grown on foundry-compatible Si (001) substrates with CW, RT operation. Despite a moderate increase in $J_{th}$ caused by threading dislocations, our visible lasers on Si (001) compare favorably with earlier-reported devices based on similar active regions grown on GaAs (001). Low-threshold, monolithically integrated visible lasers on Si can serve as an important low-cost enabler for visible optoelectronics applications ranging from quantum information [4] to near-eye displays [30].

All lasers were grown in a Veeco Mod Gen II solid-source molecular beam epitaxy (MBE) system on GaAs (001) and GaAs/Si (001) without any intentional offcut. We grew relaxed GaAs on GaP/Si (001) templates commercially available from NAsP$_{III-V}$ GmbH using a combination of thermal cycle annealing and dislocation filtering [31] [Supplement 1]; the total thickness of the buffer layer was ~2.15 μm. The laser structure [Fig. 1(a)] includes an optical cavity consisting of 1000 nm n- and p-Al$_{0.51}$In$_{0.49}$P (AlInP, hereafter) cladding layers and a 150 nm (Al$_x$Ga$_{1-x}$)$_{0.52}$In$_{0.48}$P (x = 0.34-1) continuous graded index waveguide (GRIN-WG) layer. We grew In$_{0.48}$Ga$_{0.52}$P barrier reduction layers (BRL) between the GaAs contact and AlInP cladding layers to mitigate voltage drops resulting from band offsets [32]. The cladding, waveguide and BRLs were lattice matched to GaAs, as confirmed by high-resolution x-ray diffraction. The active region of the single QW (SQW) laser consists of a compressively strained 7 nm In$_{0.6}$Ga$_{0.4}$P QW surrounded by 50 nm, 2.1 eV Al$_{0.17}$Ga$_{0.35}$In$_{0.48}$P (AlGaInP, hereafter) spacer layers, lattice matched to GaAs. The active region of the InP multiple quantum dot (MQD) lasers utilizes a QD in a well design (QDWELL) with 3.5 monolayers (ML) InP QDs capped by a 7 nm In$_{0.6}$Ga$_{0.4}$P QW and surrounded by 16 nm AlGaInP spacer layers; the QDWELL structure was repeated 3× in the InP MQD laser. All laser structures underwent post-growth rapid thermal annealing (RTA) at 950°C for 1 s to improve the optical quality of the active region [33] prior to fabrication of uncoated, broad-area lasers [Supplement 1]. Details of the beneficial effect of RTA on both photoluminescence and laser threshold characteristics will be discussed in a future publication. Laser testing was performed under CW injection with devices sitting on a temperature-controlled stage [See Supplement 1 for further details]. Reliability studies are planned for the future, but we see no evidence of degradation over the time spent characterizing these devices.

The bright-field transmission electron microscope (BF-TEM) image in Fig. 1(b) shows an entire InP MQD laser structure grown on GaAs/Si. The striated contrast throughout the device is common for ternary and quaternary AlGaInP alloys grown by MBE and results from weak phase separation during growth [34]. The active region shows 3 layers of coherently strained InP QDs exhibiting a mottled, dark strain contrast, while the apparent absence of threading dislocations indicates that the TDD in the active region is close to or below the detection limit of $\sim 1 \times 10^7$ cm$^{-2}$. Importantly, no misfit dislocations are observed in the active region, despite the compressive strain present in both the InP QDs and In$_{0.6}$Ga$_{0.4}$P QWs. A high-angle annular dark-field scanning TEM (HAADF-STEM) image of a single QDWELL layer in Fig. 1(c) shows the composition contrast of individual InP QDs. The density of buried InP QDs, calculated using BF-TEM is > $1 \times 10^{11}$ cm$^{-2}$, consistent with our previous report [35]. The panchromatic cathodoluminescence (CL) map in Fig. 1(d) of the n-GaAs/Si virtual substrate used for the laser growth confirms the TDD value of ~ $1 \times 10^7$ cm$^{-2}$.

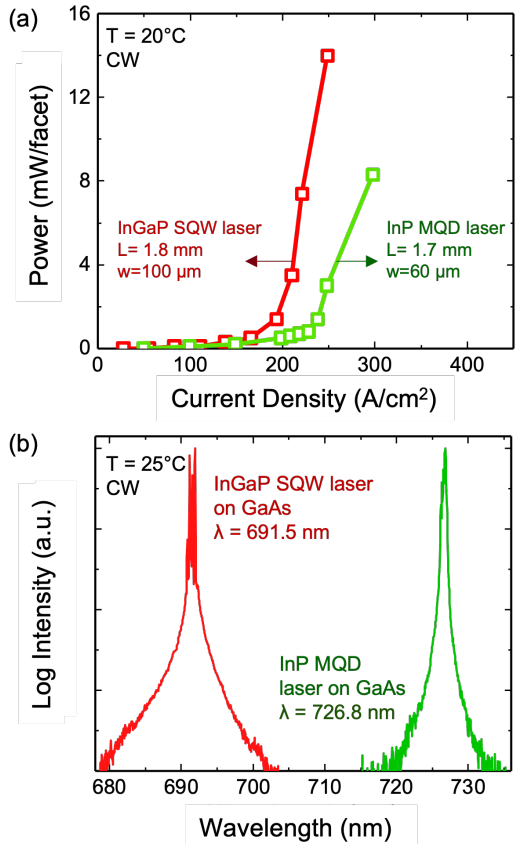

**Fig. 2.** Benchmark laser characteristics for devices grown on GaAs: (a) L-I curves for $In_{0.6}Ga_{0.4}P$ SQW (red, 1.8 mm cavity length with 100 μm ridge) and InP MQD (green, 1.7 mm cavity length with 60 μm ridge) lasers tested CW at 20°C, with $In_{0.6}Ga_{0.4}P$ SQW (InP MQD) laser exhibiting $J_{th}$ = 170 A/cm$^2$ (230 A/cm$^2$). (b) Semi-logarithmic laser spectra showing $In_{0.6}Ga_{0.4}P$ SQW laser emitting at 691.5 nm and InP MQD laser emitting at 726.8 nm with multiple modes. The spectra were collected at RT under CW operation.

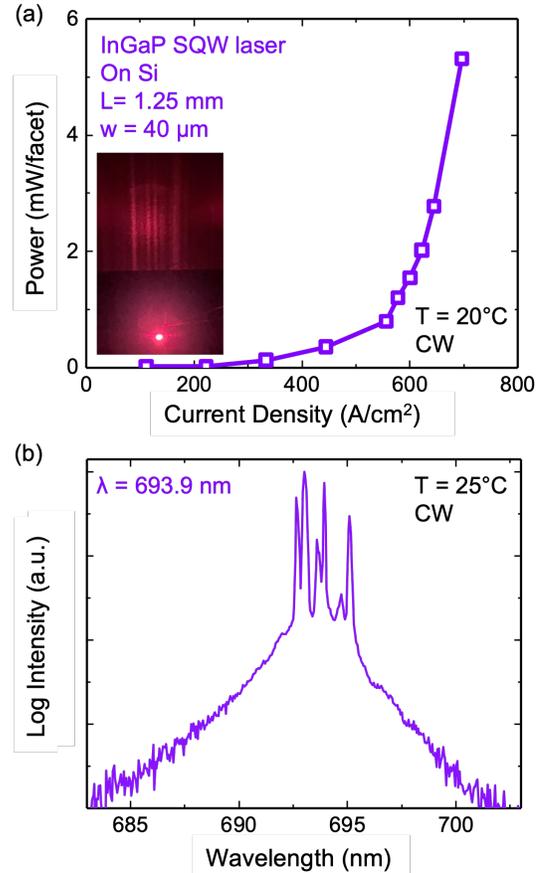

**Fig. 3.** (a) L-I curve for $In_{0.6}Ga_{0.4}P$ SQW laser on GaAs/Si (001) with a cavity length of 1.25 mm and ridge width of 40 μm operating CW at T = 20°C with $J_{th}$ = 550 A/cm$^2$. (inset) Photograph of $In_{0.6}Ga_{0.4}P$ SQW laser on Si lasing with output power > 5 mW projected on a wall ~50 cm away from device-under-test. (b) Semi-logarithmic lasing spectra of $In_{0.6}Ga_{0.4}P$ SQW laser on Si operating above threshold, collected at RT CW showing multiple mode emission centered at 693.9 nm.

A critical precursor towards demonstrating low-$J_{th}$ red lasers on Si was to develop high-performance benchmark devices on native GaAs substrates, and Fig. 2(a) shows the light intensity vs current density (L-I) characteristics of broad-area $In_{0.6}Ga_{0.4}P$ SQW and InP MQD lasers on GaAs tested under CW operation at 20°C. The $In_{0.6}Ga_{0.4}P$ SQW laser shows a low $J_{th}$ of 170 A/cm$^2$ with an output power of > 10 mW, while the InP MQD laser exhibits a slightly higher $J_{th}$ of 230 A/cm$^2$ (77 A/cm$^2$ per QDWELL layer) due to its thicker active region. Although a pulsed $J_{th}$ value of 190 A/cm$^2$ has been reported previously for InP MQD lasers, these are the lowest CW $J_{th}$ values on GaAs (001) that we are aware of. Fig. 2(b) shows that the $In_{0.6}Ga_{0.4}P$ SQW laser emits at 691.5 nm and the InP MQD laser emits in the far-red regime at 726.8 nm; both exhibit multiple transverse and longitudinal modes, as expected for broad-area lasers. The ultra-low CW $J_{th}$ of our $In_{0.6}Ga_{0.4}P$ SQW and InP MQD lasers on GaAs establishes that our material quality is at or near state-of-the-art values and enables us to observe the performance of our lasers on Si without the deleterious point defects that have been reported in MBE-grown phosphides [36].

$In_{0.6}Ga_{0.4}P$ SQW lasers on Si (001) exhibit a CW $J_{th}$ of 550 A/cm$^2$ [Fig. 3(a)], 3× lower than previously reported pulsed devices on offcut Si [28]. The inset of Fig. 3(a) shows a photograph of the laser (cavity length = 1.25 mm, ridge width = 40 μm) located at the bottom/foreground operating at ~ 5 mW output power with the far-field pattern projected onto a wall, ~50 cm from the device-under-test; the dark/bright vertical stripes in the far field result from diffraction effects in the elliptical emission pattern. The SQW laser on Si (001) emits with multiple modes centered at 693.9 nm as shown in Fig. 3(b), slightly redshifted compared to the laser grown on GaAs. The redshift could be attributed to tensile strain arising from the thermal mismatch between III-V layers and Si [37]. Despite a TDD of $1 \times 10^7$ cm$^{-2}$, $In_{0.6}Ga_{0.4}P$ SQW lasers show only a moderate increase of 3.2× in $J_{th}$ compared to SQW lasers on GaAs. Besides the impact of threading dislocations, the shorter cavity length (1.25 vs 1.8 mm) of the SQW laser on Si and the narrower ridge width (40 μm vs 100 μm) could be responsible for the increase in $J_{th}$. The relatively low $J_{th}$ of our $In_{0.6}Ga_{0.4}P$ SQW lasers on Si is noteworthy considering that earlier work used GaAs/Si with a much lower TDD of 2×10$^6$ cm$^{-2}$ [28]. We believe that non-radiative recombination at point defects is the most likely reason for the high $J_{th}$ values observed by Kwon et al. on both GaAs and Si, which in turn dominates the effects of threading dislocations. For comparison, In$_{0.15}$Ga$_{0.85}$As QW lasers on GaAs/Si with emission at ∼1 μm and TDD ∼ 1×10$^8$ cm$^{-2}$ exhibited pulsed $J_{th}$ of 5.6 kA/cm$^2$, ∼60× higher than their counterparts grown on GaAs (001) [37]. Our In$_{0.6}$Ga$_{0.4}$P SQW lasers on Si appear to show a comparatively higher degree of tolerance to threading dislocations, which may result from the low carrier diffusivity in phosphides compared to arsenides [38]. In contrast, Ga$_x$In$_{1-x}$As$_y$Sb$_{1-y}$ QW lasers grown on GaSb on Si emitting at 2.3 μm with a TDD = 1.4×10$^8$ cm$^{-2}$ show only ∼2× increase in $J_{th}$ compared to lasers grown on GaSb [39]. Further study is needed to better understand the complex interplay of bandgap energy and composition on the dislocation-tolerance of III-V lasers.

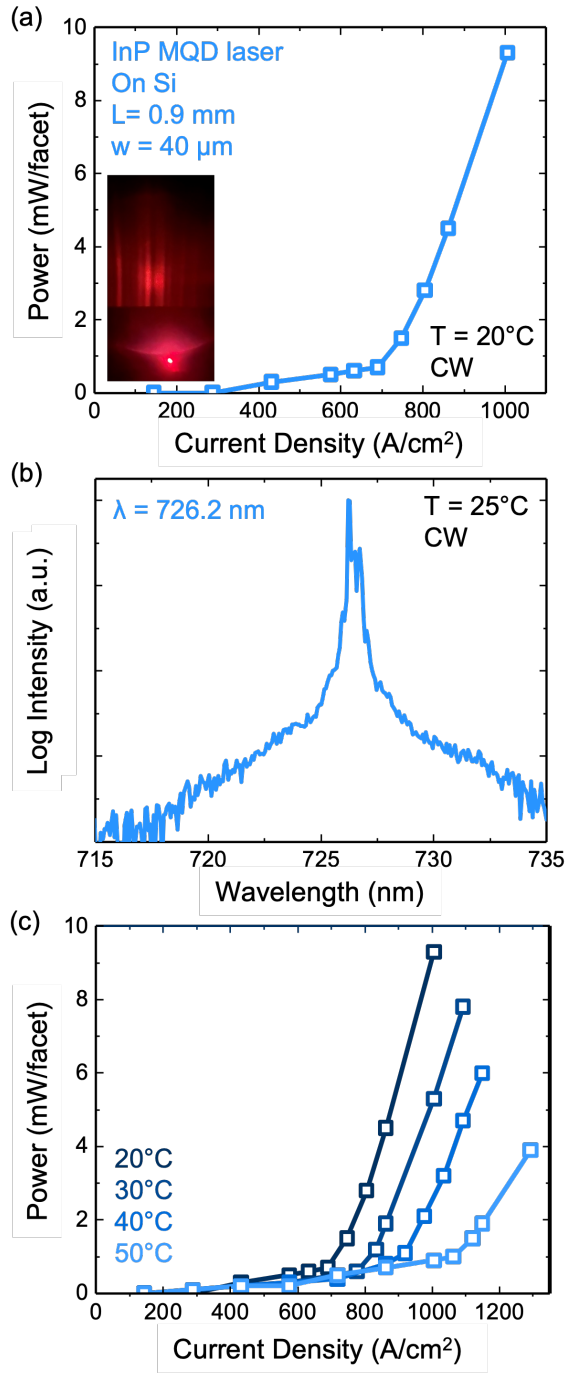

**Fig. 4.** (a) L-I curve for InP MQD laser on GaAs/Si (001) with a cavity length of 0.9 mm and ridge width of 40 μm operating CW at T = 20°C with $J_{th}$ = 690 A/cm$^2$. (inset) Photograph of InP MQD laser on Si lasing with output power∼5 mW projected on a wall ∼50 cm away from device-under-test. (b) Semi-logarithmic lasing spectra of InP MQD laser on Si operating above threshold, collected at RT under CW operation showing multiple mode emission centered in the far-red regime at 726.2 nm. (c) L-I curves for InP MQD laser on Si operating CW at elevated temperature of 20°C to 50°C showing an increase in $J_{th}$ with characteristic temperature of 65 K.

Fig. 4(a) shows the RT, CW L-I characteristics of the first electrically injected InP MQD laser on GaAs/Si. $J_{th}$ of this laser is 690 A/cm$^2$ (230 A/cm$^2$ per QDWELL layer), and the inset of Fig. 4 (a) shows a photograph of the InP MQD laser (cavity length = 0.9 mm, ridge width = 40 μm) operating at ∼ 5 mW output. The InP MQD laser on Si emits with multiple modes centered at 726.2 nm [Fig. 4(b)], nearly identical to our MQD lasers on GaAs. Unlike the slightly different wavelengths of the SQW lasers described above, here we attribute the lack of redshift to minor differences in QD growth on GaAs vs GaAs/Si. Fig.4(c) shows that $J_{th}$ of the InP MQD laser on Si increases from 690 A/cm$^2$ at 20°C to 1063 A/cm$^2$ at 50°C. We extracted a characteristic temperature $T_o$ of 65 K for InP MQD lasers on Si, which is lower than the value of 88 K for lasers on GaAs [Supplemental 1]. The lower $T_o$ on Si indicates the need for improved heat dissipation in the active region and further reduction of TDD [40].

The InP MQD laser on GaAs/Si shows a $J_{th}$ increase of 3× compared to its counterpart grown on GaAs, which is comparable to the 2× increase typically seen in InAs MQD lasers on Si [10]. Like QW lasers on Si, a part of the increase in $J_{th}$ could be attributed to the shorter cavity length and narrower ridges. But based on previous PL studies where InP QDs showed similar intensity on both GaAs and GaAs/Si [29, 35], we would have expected that the carrier confinement offered by the QDs would confer some $J_{th}$ advantage for laser operation over the QWs. The high-level carrier injection inherent to laser operation may partly explain the qualitative discrepancy between the PL (taken at very low-level injection) and laser results. Future studies with optimized device design and processing could help further unveil the effects of threading dislocations on visible QW and QD lasers grown on Si.

Fig. 5(a) shows that we achieved low-$J_{th}$ operation for red and far-red lasers on both GaAs and GaAs/Si substrates. Despite a $J_{th}$ increase of ∼3× caused by threading dislocations, both In$_{0.6}$Ga$_{0.4}$P SQW and InP MQD lasers on Si show comparable $J_{th}$ to previously published red and far-red lasers grown on GaAs (Supplement 1). We believe that the use of a GRIN design for optical and electrical confinement, reduction of non-radiative recombination centers using RTA, and the inherent low diffusivity of carriers in phosphides are among the key factors for our low-$J_{th}$ lasers on Si substrates. Fig. 5(b) shows the spectra of In$_{0.6}$Ga$_{0.4}$P SQW and InP MQD lasers on GaAs and Si spanning from 680-730 nm. The emission wavelength range can be tailored to a wide range of applications by utilizing tensile-strained In$_x$Ga$_{1-x}$P QWs [41] for shorter wavelength and alloying InP QDs with arsenic for longer wavelength emission [42].

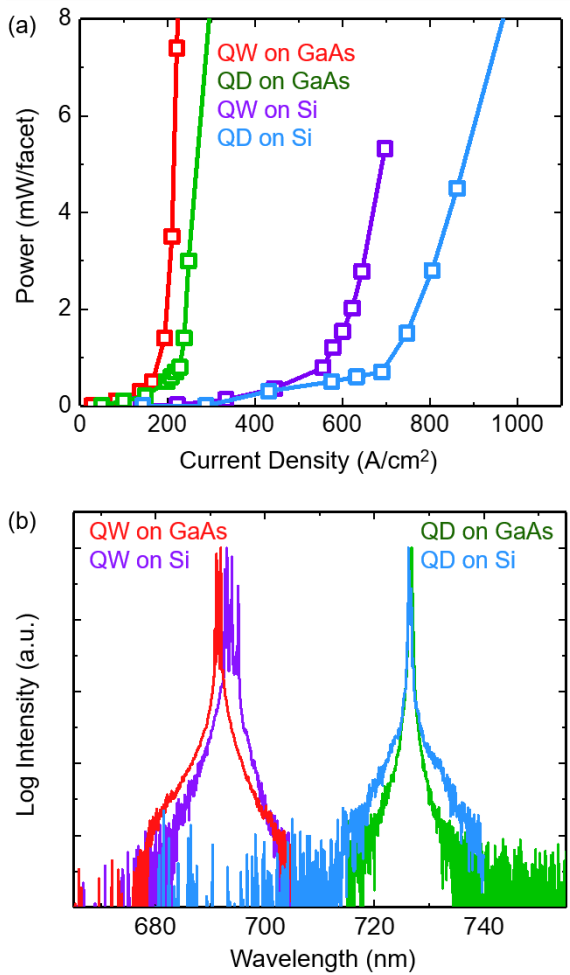

**Fig. 5.** (a) Comparison of L-I characteristics of $In_{0.6}Ga_{0.4}P$ SQW and InP MQD lasers on GaAs and GaAs/Si with both SQW and MQD lasers on Si showing a $J_{th}$ increase of ~3× compared to lasers on GaAs. (b) Spectra of $In_{0.6}Ga_{0.4}P$ SQW and InP MQD lasers grown on GaAs and GaAs/Si in semi-logarithmic scale.

The performance of both the SQW and MQD lasers could be further improved by reducing the TDD of GaAs/Si virtual substrates [13] and exploring p-modulation doping in the QD active regions [43]. Future work will also aim to lower the threshold voltage of the devices by reduction of majority carrier barriers and optimization of contact resistances [44]. For example, the $In_{0.6}Ga_{0.4}P$ SQW lasers on both GaAs and Si operate with a threshold voltage of 2.7 V, which is higher than expected for a laser emitting at a photon energy of ~1.8 eV; previously published lasers in this wavelength range operated at voltages of 2.1-2.3 V [33, 45], indicating room for improvement. In addition to facet coating, we believe that lower $J_{th}$ operation of $In_{0.6}Ga_{0.4}P$ QW and InP MQD lasers on Si can be achieved with systematic improvements in facet formation [18], as well as longer cavity lengths [33]. Future work will also aim towards testing the extrapolated operation lifetime of $In_{0.6}Ga_{0.4}P$ SQW and InP MQD lasers on Si and understanding their degradation mechanisms.

In conclusion, we demonstrated the first RT, CW, electrically injected red $In_{0.6}Ga_{0.4}P$ SQW and far-red InP MQD lasers on Si (001) with respective $J_{th}$ values of 550 A/cm$^2$ and 690 A/cm$^2$. This study indicates that the effect of dislocations on phosphide- and arsenide-based lasers on Si differ significantly, with arsenides showing stronger benefits in $J_{th}$ by switching from a QW to a QD active region. III-V lasers based on diverse active region designs, compositions, and bandgap energies can all behave differently when grown on Si, and future studies will undoubtedly lead to deeper insights on these differences. Phosphide-based QW and QD lasers offer high performance over a wide range of wavelengths from 630-800 nm, and this work establishes that such lasers can be grown on Si (001). Combined with $SiN_x$ waveguides, such short-wavelength lasers open the intriguing possibility of direct integration with highly sensitive Si photodetectors [46], circumventing the escalated dark current of epitaxial Ge/Si detectors [47]. Epitaxial QD and QW lasers emitting at 1.3-2.3 μm are becoming increasingly established, and this work takes a vital first step towards integration of red visible lasers that will allow integrated photonics to expand its impact into areas such as on-chip biosensing [48] and quantum computing [9].

**FUNDING**

We gratefully acknowledge funding from MIT Lincoln Laboratory under the program, "Heteroepitaxial III-V/SiN$_x$ Integrated Photonics (HIP)". R.D.H and B.D.L were supported by NASA Space Technology Research Fellowships under grant numbers 80NSSC18K1171 and 80NSSC19K1174, respectively.

**ACKNOWLEDGEMENTS**

We thank Chris Heidelberger, Reuel Swint, and Paul Juodawlkis for helpful discussions and assistance. We also thank Katherine Lakomy and Prof. Kent Choquette for help with initial pulsed testing of the lasers.

**DISCLOSURE**

The authors declare no conflicts of interest.

**DATA AVAILABILITY**

Data underlying the results presented in this paper are not publicly available at this time but may be obtained from the authors upon reasonable request.

**SUPPLEMENTAL DOCUMENT**

See Supplement 1 for supporting content.